\documentclass[12pt]{iopart}

\usepackage{iopams}

\usepackage{graphicx}% Include figure files

\begin{document}
\title[On the nature of large-scale defect accumulations in
Czochralski-grown silicon]{On the nature of large-scale defect accumulations in Czochralski-grown silicon}

\author{V~P~Kalinushkin,
A~N~Buzynin, V~A~Yuryev,
O~V~Astafiev\  and D~I~Murin}

\address{General Physics Institute of the Russian Academy of Sciences,
38, Vavilov Street, Moscow, GSP--1, 117942, Russia}
%\affil{General Physics Institute of the Russian Academy of Sciences,
%38, Vavilov Street, Moscow, GSP--1, 117942, Russia}

\begin{abstract}
Czochralski-grown boron-doped silicon crystals were studied by the techniques
of the low-angle mid-IR-light scattering and electron-beam-induced current.
The large-scale accumulations of electrically-active impurities detected in
this material were found to be different in their nature and formation
mechanisms from the well-known impurity clouds in a float zone-grown silicon.
A classification of the large-scale impurity accumulations in CZ Si:B
is made and point centers constituting them are analyzed in this paper.
A model of the large-scale impurity accumulations in CZ-grown Si:B is also
proposed.
\end{abstract}

\section{Introduction}
The detection of the large-scale impurity accumulations (LSIAs) with
the sizes ranged from several to several tens $\mu$m in CZ Si by means
of the low-angle light scatter (LALS) \cite{4}
was reported for the first time in
Ref.\,\cite{1}. It was supposed in that work that LSIAs are analogous in their
nature to the oxygen and carbon clouds observed in FZ Si in Ref.\,\cite{2}.
It was shown, however, as a result of the research of Si crystals grown at
variable growth rate done by LALS and EBIC that most of LSIAs in CZ Si
have a shape close to cylindrical \cite{3} which contradict the cloud
model \cite{1}. In the present work, an attempt is made to select different
in their nature types of LSIAs in CZ Si and an information about their
parameters as well as the influence of different thermal treatments on them
is given.

\section{Experimental details}
Industrial 76 and 100-nm substrates of CZ Si:B grown in the $<$100$>$
and $<$111$>$ directions\,---\,$\varrho  \sim$ several $\Omega$\,cm\,---\,were
studied in this work. The oxygen concentration in them ranged from
6\,$\times$\,10$^{17}$ to 10$^{18}$~cm$^{-3}$, the carbon concentration was less
than 10$^{16}$\,cm$^{-3}$. The thickness of 76-mm substrates was 380\,$\mu$m,
that of 100-mm ones was 500\,$\mu$m.

The investigation was carried out by LALS and EBIC. CO- and CO$_2$-lasers
oscillating at the wavelength of 5.4 and 10.6~$\mu$m, respectively,
were used in LALS to select the scattering by free carrier accumulations
\cite{5}. To determine the activation energies ($\Delta E$) of the centers
constituting LSIAs, the temperature dependances of LALS were investigated
in the range from 85 to 300\,K \cite{6}. A shape of LSIAs was determined from the
dependances of the LALS diagrams on the sample orientation with respect to
the detection plane. The plasma etching of the sample surface
in a special regime before the Schottky barrier creation greatly increased the
sensitivity of EBIC to electrically-active defects in crystals \cite{7}.

In the experiments on annealings, wafers were cut into four sections. One
of them was not treated, the others were subjected to either isothermal
processes at 600 or 800$^{\circ}$C for 24, 48 and 120~h, respectively, or
high-temperature treatments at 965, 1100, 1150, 1200 and 1250$^{\circ}$C
for several tens minutes. The treatments at $T > 1200^{\circ}$C resulted in
the formation of a large amount of defects of structure which were
revealed by the selective etching (SE). The substrates grown in the $<$100$>$
direction were subjected to both the former and the latter treatments, while
those grown in the $<$111$>$ direction were treated only in the latter way.

\section{Results}
\subsection{Initial samples}
\begin{figure}[t] %1
\vspace*{0.5mm}
\hspace*{26.5mm}
\includegraphics[scale=1]{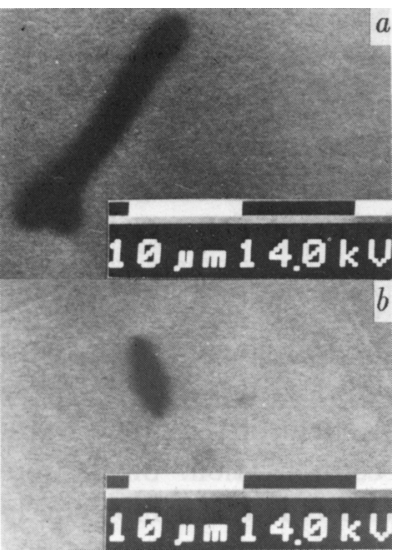}
\includegraphics[scale=1]{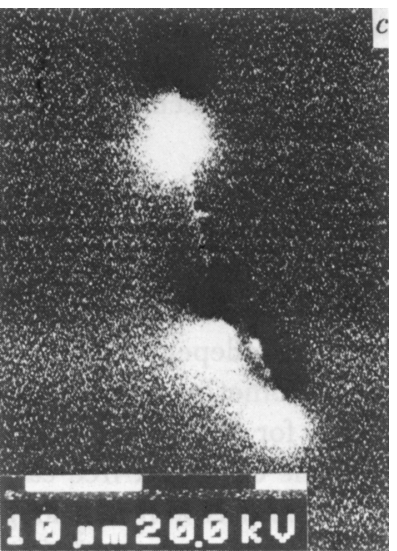}
\includegraphics[scale=1]{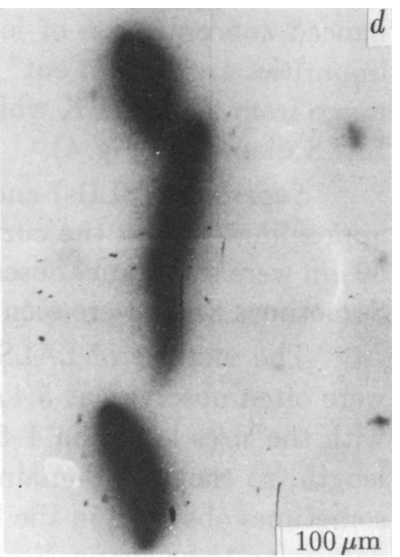}
\caption{EBIC microphotographs of as-grown CZ Si:B:
cylindrical $(a,b)$, spherical $(c)$ and superlarge $(d)$
defects.}\label{f1}
\end{figure}

Fig.\,\ref{f1} demonstrates the EBIC images of defects. The samples
contain many non-uniformities with the sizes from
several to several tens $\mu$m\,---\,mainly cylindrical. In
addition, spherical defects are also seen.

\begin{figure}[t]%2,3,4
 \vspace*{0.5mm}
\hspace*{15mm}
\includegraphics[scale=2.5]{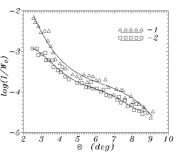}
\includegraphics[scale=2.5]{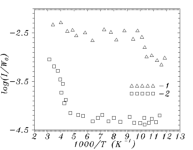}
\includegraphics[scale=2.5]{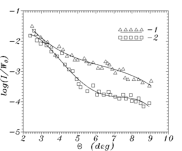}
\caption{LALS diagrams for initial CZ Si:B, orientation
with respect to detection plane (deg): 0 (1), 90 (2).}\label{f2}
\caption{LALS temperature dependances for cylindrical (1) and
spherical (2) defects in initial CZ Si:B.}\label{f3}
\caption{LALS diagrams at different temperatures (K):
300 (1), 110 (2).}\label{f4}
\end{figure}

{\it Cylindrical defects} (CDs).
There are sections of LALS diagrams, the shape of
which is dependent on the sample orientation to the detection plane
(Fig.\,\ref{f2}, $\theta < 4.5^{\circ}$). These sections are well
fitted with the curves of scattering by cylinders \cite{4} with the diameters
from 3--4 to 8--10~$\mu$m and length from 15 to 40~$\mu$m depending on a
sample. They predominantly oriented along the $<$110$>$ direction.
It is seen from the EBIC patterns that CDs have rather elliptical or
curved-cylindrical shape (Fig.\,\ref{f1}\,$(a,b)$). We assume the CDs
revealed by EBIC and LALS to be the same defects
similar to CDs observed in Ref.\,\cite{3}.

The concentration of CDs\,---\,the most usual defects in CZ Si:B\,---\,estimated
from EBIC ranged from 10$^6$ to 10$^7$~cm$^{-3}$. We could not find
a dependance of the CD concentration on oxygen concentration, growth
direction, ingot diameter or location on a wafer. Nonetheless we found
their concentration to vary within a wafer as well as in different wafers.

LALS at 10.6 and 5.4-$\mu$m wavelength showed CDs to be domains of the enhanced
free carrier concentration \cite{5}. Using the CD concentrations from
EBIC, the variations of the dielectric function ($\Delta \varepsilon$) and
the maximum free carrier concentration in them  ($\Delta n_{max}$) were
evaluated \cite{4}: they are (1--4)\,$\times$\,10$^{-4}$ and
(3--10)\,$\times$\,10$^{15}$\,cm$^{-3}$, respectively.
CDs occupy less than 3\,\% of crystal volume and the total amount of
impurities contained  in them ($N_i$) is not greater than
3\,$\times$\,10$^{14}$\,cm$^{-3}$.

LALS temperature dependances showed for CDs a small (2--3 times)
drop of the scattering intensity ($I_{sc}$) at about 90\,K
(Fig.\,\ref{f3}, curve 1). This allows us to state that LALS by CDs
at 300\,K at 10.6\,$\mu$m is controlled by the centers with
$\Delta E \approx$\,40--60\,meV containing in CDs. Naturally
another defects, such as deep or compensating centers as well as precipitates,
inclusions and  structural imperfections can also be contained in CDs.

{\it Spherical defects} (SDs). Besides the above sections of LALS diagrams,
those independent of the sample orientation were also observed
(Fig.\,\ref{f2}, $\theta > 4.5^{\circ}$). These sections are well
fitted with the curves of light scattering by spherical defects with
the Gaussian profile of $\varepsilon$ \cite{4}
and sizes from 5--8 to 20~$\mu$m. Such defects are also seen in the EBIC
pictures (Fig.\,\ref{f1}\,$(c)$). Their concentration is usually about
10$^5$\,cm$^{-3}$, $\Delta \varepsilon \approx$\,(1--3)\,$\times$\,10$^{-3}$,
$\Delta n_{max} \approx$\,(3--9)\,$\times$\,10$^{16}$\,cm$^{-3}$ \cite{4}.
SDs occupy less than 0.04\,\% of the crystal volume,
$N_i  {_{^{\sim}}^{_<}}\,4\,\times\,10^{13}$\,cm$^{-3}$.

LALS temperature dependances showed SDs, like CDs, to be domains with
enhanced concentration of ionized at 300\,K impurities with
$\Delta E \approx$\,120--160\,meV. These impurities are ``frozen out'' at
about 250\,K, so SD-related scattering is smothered in the range from
90 to 250\,K which enables the accurate selection of CD-related scattering in
LALS diagrams (Fig.\,\ref{f4}).

{\em Superlarge {\em (SLDs) and {\em small} (SmDs)} defects}. The sections
of LALS diagrams well approximated with the curves of light scattering by
defects with the sizes greater than 50\,$\mu$m were sometimes observed.
These defects appeared to have an asymmetrical shape. Sometimes SLDs
were seen in the EBIC photographs (Fig.\,\ref{f1}\,$(d)$).

The sections of LALS diagrams independent of the scattering angle
(``plateaux'') were often observed at 5.4\,$\mu$m (and sometimes at
10.6\,$\mu$m) which correspond to defects with the sizes less than
4--5\,$\mu$m. $I_{sc}$ for SmDs was also independent of the probe
wavelength, so they are  domains with the enhanced free carrier concentration.

SmDs were sometimes observed in the EBIC picture as well. Although we
could not unambiguously determine their shape, SmDs seem to be very small
CDs and SDs rather than a separate class of defects. This was verified
by LALS temperature dependances: SmDs were ``frozen out'' at 90\,K
when CDs predominated and at 250\,K if SDs predominated.

\subsection{Annealed samples}
The LALS diagrams and EBIC pictures for the annealed crystals did not
differ in general features from those for the as-grown samples. The
following peculiarities may be emphasized.

1. In crystals annealed in the temperature range of 600--1100$^{\circ}$C,
the light scatter by SDs was greatly (but not completely) suppressed,
and CDs and SmDs predominated. EBIC showed mainly CDs and SmDs
too.

\begin{figure}[t]
 \vspace*{0.5mm}
\hspace*{26.5mm}
\includegraphics[scale=1]{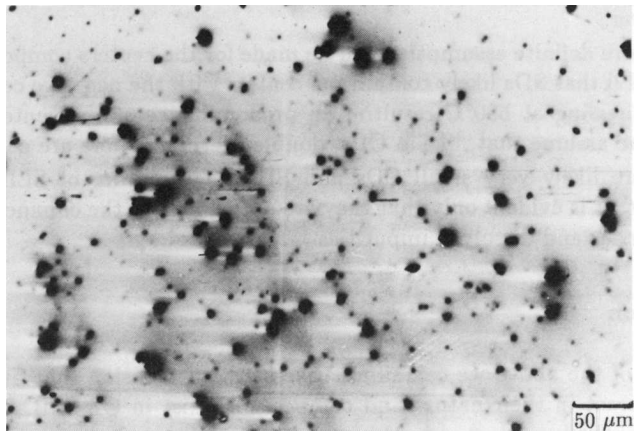}
\caption{Typical EBIC microphotograph of defects in CZ Si:B annealed
at $T\,>\,1100^{\circ}$C.}\label{f5}
\end{figure}

2. Annealing at $T > 1100^{\circ}$C resulted in predominance of
SD-related scattering and general growth of $I_{sc}$. A great number
of SDs was observed by EBIC (Fig.\,\ref{f5}).

3. After annealing at $T > 1200^{\circ}$C, the centers with the same $\Delta E$
as in the as-grown samples composed LSIAs.

4. Annealing at 800$^{\circ}$C and short (up to 48\,h) treatment at  600$^{\circ}$C
did not change $\Delta E$ of the centers composing CDs and SDs. Longer
treatment at 600$^{\circ}$C resulted in prevailing of the centers with
$\Delta E \approx$\,70--90\,meV in CDs and SDs.

5. After 120-h annealing at 600 and 800$^{\circ}$C, SLDs became more habitual
than in the as-grown samples. The centers with $\Delta E \approx$\,130--170\,mev
were contained in SLDs.

\section{Discussion}
It is difficult now to determine the nature of LSIAs in CZ Si:B unambiguously.
Ii is clear, however, that CDs are domains with the enhanced free carrier
concentration caused by point centers with $\Delta E \approx$\,40--60\,meV.
CDs look like the defects observed by EBIC after oxidizing annealings
\cite{8}. Our research gave an evidence to the presence of these defects
in initial crystals, moreover the CD-related sections of LALS diagrams
do not change after high-temperature annealings. In our opinion, however,
CDs are the impurity atmospheres (IAs) around defects-precursors, e.g.
stacking faults (SFs). Remark that some authors connect the contrast of
EBIC patterns with the formation of precipitate colonies around SFs \cite{8}.
As these colonies may have no influence on the free carrier concentration
in IAs, the following scenario might be proposed.

In initial wafers, the precipitate concentration in IAs is low.
At the same time, the dissolved
ionized impurity concentration  is high enough but insufficient for
the precipitate formation, hence $I_{sc}$ is high and the recombination
contrast in EBIC is  low. During high-temperature annealing,
precipitate colonies arise in CDs but the dissolved impurity concentration
and $\Delta n$ change weakly (e.g. because the impurity concentration
in CDs in the as-grown samples was close to the saturation limit or due to
the growth of the compensation degree). The recombination contrast will
have grown and $I_{sc}$ will have changed weakly and randomly.

From the other hand, the enhanced EBIC contrast may be caused by the
specificity of the sample preparation. Two variants are possible.
The centers enhancing the EBIC contrast may arise as a result of
the plasma etching applied. Alternatively, an ``exhaustion'' of CDs may be
a result of the chemical etching usually applied for sample preparation
for EBIC.

Thus, the hypothesis according to which CDs are IAs around SFs does not meet
contradictions. As to the point defects composing CDs, we suppose them to be
the ``new'' thermal donors \cite{9} whose $\Delta E$ is close to the estimates
made\,\footnote{Although, some alternatives exist \cite{10}, and B and [Cu--O]
are among them.}. The influence of 600$^{\circ}$C annealing on $\Delta E$
of the centers composing CDs indirectly verifies the assumption\,\footnote{The
growth of ionization energy of ``new'' thermal donors as a result
of long-term annealing at 650$^{\circ}$C was reported in \cite{11}.}. New
experiments are required to obtain more evidences to the model
proposed, though.

SDs also are domains with the enhanced dissolved impurity and free carrier
concentrations. We assume SDs to be IAs around defects of structure (e.g.
precipitates). This hypothesis is confirmed by the growth of the SD
concentration during the high-temperature annealings and correlation
with the appearance of structural defects revealed by SE. This assumption
also have no sufficient evidences\,\footnote{Some alternatives
to the assumption\,---\,the cloud models\,---\,are discussed in \cite{10}.}
and require an additional research, however.

Some more definite assumption can be made for the centers composing SDs.
It was supposed in \cite{12} that SDs likely contain the centers with
the negative correlation energy. Regarding annealing at
600$^{\circ}$C resulting in
predominance of the centers with changed $\Delta E$ in SDs, we assume
that, like in CDs, double thermal donors are contained
in SDs\,\footnote{Possible
alternatives to these centers are discussed in \cite{10}, they are
B$_i$, [O--V], [C$_i$--C$_s$], {\em etc}.}.

SmDs are likely very small CDs and SDs. The nature of SLDs is hard to be
discussed now. It is evident only that they are domains with the enhanced
concentration of the free carrier and dissolved impurities.

\section{Conclusion}
On the basis of the above we can summarize in the conclusion that
at least two types of LSIAs different in their nature and composition exist in
CZ Si:B. Their parameters determined for the investigated in this work group of crystals
are rather typical for the industrial Si:B with the specific
resistivity of several $\Omega$\,cm. Some additional details of this
research can be found in Ref.\,\cite{10}.\\

\end{document}